\PassOptionsToPackage{bookmarks=false}{hyperref}
\documentclass[conference]{IEEEtran}
\usepackage{balance}
\usepackage{xspace}
\usepackage{hyperref}
\usepackage{listings}
\usepackage{tcolorbox}
\usepackage{amsmath,amssymb,amsfonts}
\usepackage{algorithmic}
\usepackage{graphicx}
\usepackage{textcomp}
\usepackage{xcolor}
\usepackage{wasysym}

\usepackage{booktabs}
\usepackage{xcolor, colortbl}
\usepackage{xspace}

\usepackage{multirow}
\usepackage{inconsolata}

\def\BibTeX{{\rm B\kern-.05em{\sc i\kern-.025em b}\kern-.08em
    T\kern-.1667em\lower.7ex\hbox{E}\kern-.125emX}}
\begin{document}

\title{Autonomous Large Language Model Agents Enabling Intent-Driven Mobile GUI Testing\\}

\author{\IEEEauthorblockN{Juyeon Yoon}
\IEEEauthorblockA{
\textit{KAIST}\\
juyeon.yoon@kaist.ac.kr}
\and
\IEEEauthorblockN{Robert Feldt}
\IEEEauthorblockA{
\textit{Chalmers University of Technology}\\
robert.feldt@chalmers.se}
\and
\IEEEauthorblockN{Shin Yoo}
\IEEEauthorblockA{
\textit{KAIST}\\
shin.yoo@kaist.ac.kr}
}

\definecolor{ao}{rgb}{0.0, 0.5, 0.0}
\newcommand{\fixme}[2][red]{\textcolor{#1}{FIXME: #2}}
\newcommand{\addcite}[2][orange]{\textcolor{#1}{ADDCITE: #2}}
\newcommand{\jy}[2][ao]{\textcolor{#1}{JY: #2}}

\newcommand{\name}{\textsc{DroidAgent}\xspace}

\maketitle

\begin{abstract}
GUI testing checks if a software system behaves as expected when users interact with its graphical interface, e.g., testing specific functionality or validating relevant use case scenarios.
Currently, deciding what to test at this high level is a manual task since automated GUI testing tools target lower level adequacy metrics such as structural code coverage or activity coverage.
We propose \name, an autonomous GUI testing agent for Android, for semantic, intent-driven automation of GUI testing.
It is based on Large Language Models and support mechanisms such as long- and short-term memory. %
Given an Android app, \name sets relevant task goals
and subsequently tries to achieve them by interacting with the app. 
Our empirical evaluation of \name using 15 apps from the Themis benchmark shows that it can set up and perform realistic tasks, with a higher level of autonomy.
For example, when testing a messaging app, \name created a second account and added a first account as a friend, testing a realistic use case, without human intervention.
On average, \name achieved 61\% activity coverage, compared to 51\% for current state-of-the-art GUI testing techniques. 
Further, manual analysis shows that 317 out of the 374 autonomously created tasks are realistic and relevant to app functionalities, and also that \name 
interacts deeply with the apps and covers more features.
\end{abstract}


\section{Introduction}

Testing mobile applications at the GUI level to ensure their quality in terms of 
functionality and usability is a critical part of app development. 
However, it also remains a costly task due to the ever-growing complexity of 
applications and inherent issues about the Android ecosystem such as rapid 
platform evolution and device fragmentation. 

To address the challenges in GUI testing, there has been a large amount of 
research efforts~\cite{li2019humanoid, su2017guided, mao2016sapienz} to 
automate the various aspects of mobile GUI testing. Most of the existing 
techniques focus on exploring the GUI states of a given app as much as 
possible. For example, DroidBot~\cite{li2017droidbot} adopts a greedy 
exploration policy that prioritises unexplored widgets as a next exploration 
target. Humanoid~\cite{li2019humanoid} adopt deep learning in hope to mimic 
human-like exploration paths by utilising pretrained models to weigh probable 
GUI actions, aligning them with frequently performed actions based on real 
human traces. More recently, reinforcement learning has been applied to GUI 
testing to effectively discover novel GUI states using curiosity-based reward 
functions~\cite{zheng2021automatic, pan2020reinforcement, zhao2022dinodroid}. While these approaches have revealed 
actual bugs in Android apps, we also note that their objectives remain 
exploration of app structures, typically quantified using activity or widget coverage,
i.e., the number of Android activities (screens) or widgets that have been covered by 
the automated testing technique.

However, a recent empirical study of Android developers~\cite{linares2017developers} reports that the primary test design strategy adopted by 
Android developers is to follow the usage model of the apps. Developers 
overwhelmingly prefer test cases that target individual features and use cases, 
which are higher level test objectives compared to activity coverage, 
a structural testing objective. The emphasis on more semantic test objectives 
can also be seen when developers are questioned about the format of 
automatically generated test cases that they ideally want. Surprisingly, the 
most popular choice is \emph{natural language} rather than any test API 
scripts. Further, developers want expected outputs, as well as steps of use 
cases and specific app features, in the automatically generated tests. These 
results reveal a gap between what is being offered by automated Android testing 
techniques, and what developers want in test automation in Android.

This paper presents \name\footnote{\name is publicly available from \url{https://github.com/testing-agent/droidagent}} with the aim of bringing the level of Android GUI 
testing automation closer to developer preferences and expectations. Instead of going after 
structural testing goals such as higher activity coverage, \name automatically 
comes up with natural language descriptions of specific tasks that can be 
achieved using the given App Under Testing (AUT), and subsequently tries to 
interact with the GUI of the AUT with specific \emph{intent} to achieve those 
tasks. If successful, 
\name will produce a GUI test case script that can achieve the specific task, 
leaving the developer with both a natural language task description as well as 
executable test scripts. To the best of our knowledge, \name is the first 
Android GUI testing technique that can automatically generate high level 
testing scenarios based on sequences of identified tasks.

\name achieves this by using multiple Large Language Model (LLM) instances
that interact and externally act as an autonomous agent. LLMs have been used 
to automate Android GUI testing before, but either in more limited contexts, 
or with much less autonomy than \name. For example, Liu et al.~\cite{liu2023fill} 
proposed an approach to generate appropriate text inputs for a given GUI widget 
by prompting an LLM with textual descriptions of the current GUI state, while Wen et 
al.~\cite{wen2023droidbot} extended DroidBot~\cite{li2017droidbot} to generate 
a sequence of GUI actions from a given textual description of task. Lately, 
GPTDroid~\cite{liu2023chatting} showed that, given a summary of past 
exploration and descriptions of current GUI state, LLMs can choose a human-like 
next event to continue the exploration. Unlike existing approaches, \name sets 
its own testing goals autonomously, and can coherently follow long-term plans 
it has generated in order to accomplish those tasks. The autonomy of \name is 
inspired by the work on LLM-based cognitive architecture~\cite{park2023generative}
as outlined for software testing in Feldt et al.~\cite{feldt2023towards}.

We have empirically evaluated \name, using 15 apps from Themis benchmark~
\cite{su2021benchmarking}, against four baselines: two traditional GUI state exploration 
techniques, DroidBot~\cite{li2017droidbot} and Humanoid~\cite{li2019humanoid}, 
an LLM-based GUI state exploration technique, GPTDroid~\cite{liu2023chatting}, 
and a random GUI testing technique, Monkey~\cite{monkey}. \name has 
automatically generated 374 unique tasks for 15 studied apps: a manual 
assessment shows that 85\% of generated tasks are relevant and viable, while 
59\% are successfully accomplished by \name. While trying to achieve these 
tasks, \name also reports the highest average activity coverage of 61\%, 
compared to 51\% achieved by Humanoid. Our results suggest that 
LLM-based autonomous agents can potentially automate Android GUI testing at a 
higher, more semantic level than GUI state exploration, and that this
can even improve lower-level coverage.

The technical contributions of this paper are as follows:

\begin{itemize}
\item We present \name, an autonomous Android GUI testing technique that can 
set and execute app specific tasks on its own. It produces natural 
language descriptions of tasks, and test scripts that achieve them. 

\item We empirically evaluate \name against four baseline techniques, using 
Android apps taken from a widely used Themis benchmark. Our results show that 
\name is capable of generating relevant and useful app usage tasks, which it 
subsequently accomplishes automatically by interacting with the GUI of the 
given app.

\item We provide a replication package of \name that includes its public 
implementation.
\end{itemize}

The rest of the paper is organised as follows. Section~\ref{sec:background} 
presents background information about Android testing as well as agents based 
on LLMs. Section~\ref{sec:framework} describes the internal architecture of 
\name, while Section~\ref{sec:example} presents an illustrative example of how 
\name operates when given an app. Section~\ref{sec:evaluation} describes the settings of our empirical evaluation, the results of which are reported in Section~\ref{sec:results}, while Section~\ref{sec:threats} discusses threats to validity. Finally, Section~\ref{sec:conclusion} concludes.

\section{Background}
\label{sec:background}

This section outlines some background information.

\subsection{GUI Testing on Android}\label{sec:background-gui-testing}

In Android mobile applications, users primarily interact with GUI components 
such as buttons, text fields, and menus. A core component of an Anrdoid app is called an activity, which generally implements one screen (or one set of coherent functionalities) of a given app~\cite{androidactivity}. The list of contained activities are available in the manifest file contained in any apps. 
An activity can be essentially viewed as a tree 
hierarchy, in which each node represents a GUI component, named ``widget" in 
this paper, or a container grouping related widgets. Android SDK provides tools 
and interfaces to query such views and interact with the contained widgets, 
enabling actions like button presses or inputting text in textfields. Various 
testing frameworks~\cite{espresso:android,li2017droidbot} have been built on top of ADB (Android Debug Bridge), part of the Android SDK, for automation.

\subsubsection{Inference of possible actions}

GUI testing is often formulated as a problem of choosing the best next action 
based on a specific GUI state~\cite{mao2016sapienz,zhao2019recdroid,su2017guided,gu2019practical}. The possible actions can be determined from 
actionable widgets with properties like ``clickable" or ``editable''. While the 
number of possible actions on a specific GUI state is limited due to the relatively 
small screen size of mobile devices, there can be numerous actions, especially 
when considering a list of items.

\subsubsection{GUI state description}

Humans typically perceive Android UI based on the visual appearance of views on 
screen. For language models, conveying such visual information can still be 
challenging. Previous work~\cite{liu2023fill,liu2023chatting, liu2017automatic} 
rather choose to use textual properties included in a widget (e.g., \texttt{resource\_id}, \texttt{content\_description}, and \texttt{text}), which provide a brief hint about the function of the widget. 

Specifically, Liu et al.~\cite{liu2023chatting} combined textual descriptions of all the contained widgets, and used it as a dynamic context for prompting the language model to generate a next action. However, this makes it challenging to convey the hierarchical structure of GUI states and widgets.
Meanwhile, another recent work~\cite{feng2023prompting} adopt the style of HTML documents to represent hierarchical structure of an Android view. %
However, since LLMs are also trained on large codebases, and on relevant formats,  
for \name we adopt JSON notation to describe GUI state and, thus, to represent hierarchical data.

\subsection{Autonomous Agents with Large Language Models}

LLMs have been proven effective in various tasks, including software testing automation~\cite{nass2023improving, lemieux2023codamosa,liu2023chatting, feng2023prompting, brie2023evaluating}. Yet, the primary use of LLMs has been through single, templated prompts possibly with few-shot examples of desired behavior~\cite{kang2023large}. Recent LLMs, like OpenAI's GPT-3.5 and GPT-4, can directly use external tools through function calls, making it easier for LLM-based libraries like LangChain~\cite{Langchain} and AutoGPT~\cite{AutoGPT} to support hybridising LLMs with external tools.

By including additional memory structures to overcome the LLMs' limited context lengths, autonomous agents can be built that combine both long-term planning and interaction with the use of external tools.
The memory component of an LLM-driven autonomous agent can be implemented as key-value stores where a local neural network model embeds the text (value) to a vector (key).
This embedding database is then used to find content similar to the current context, e.g., actions that have already been tried in the current state.
In \name, we use the ChromaDB library~\cite{chromadb} which implements such an embedding database for texts.

With long-term memory and the ability to use tools, autonomous agents can be built to achieve significant goals. The design of such agent architectures can be based on cognitive architectures~\cite{laird2019soar}, that was originally proposed as models of the human mind, e.g., multi-agent social simulation~\cite{park2023generative}, and 3D world exploration~\cite{wang2023voyager}, and typically also include planning and reflection.
When designing \name it is particularly important that it can recover from undesired exploration paths and that it preserves knowledge about the application as exploration continues. In the following, we describe in detail how our agent-based design achieves this.

\begin{figure*}[ht]
    \centerline{\includegraphics[width=1.0\textwidth]{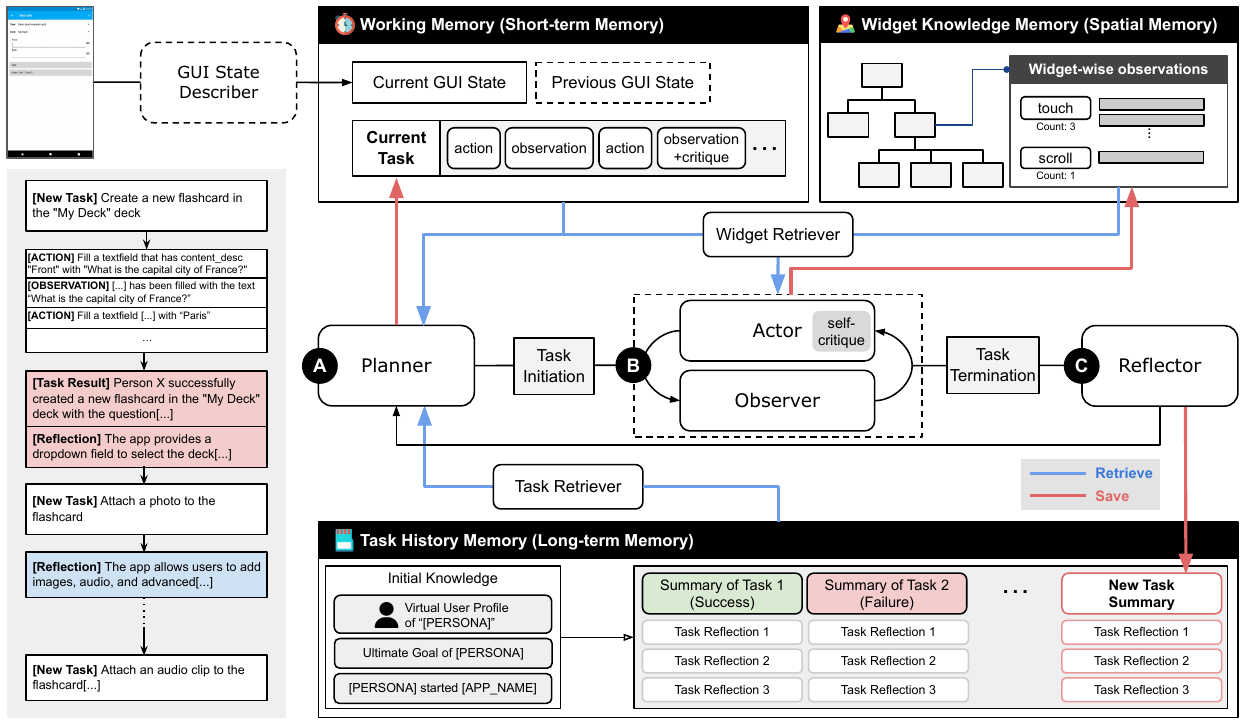}}
    \caption{Overview of \name with a task example.}
    \label{fig-architecture}
\end{figure*}

\section{Framework}
\label{sec:framework}

\name is designed with an agent-based architecture with four main LLM-based 
agents performing specific tasks: Planner, Actor, Observer, Reflector. It is 
further supported by three different memory modules (short, long, and spatial), 
with two retrieval modules that extract relevant information from memory and 
format it for use by the agents.
The Actor and Observer form an ``inner'' loop trying to perform tasks which have been planned by the Planner and which is later reflected upon by the Reflector.
Figure~\ref{fig-architecture} illustrates the role of each component and highlights the main information flow. We describe individual components of \name below; further, Section~\ref{sec:example} will provide a concrete working example.

\subsection{Task Planner}\label{sec:framework-planner}
A key part of \name is the continuous planning of high-level tasks to be 
achieved. These tasks can also directly be used to ``describe'' what 
lower-level actions, performed by other agents, actually mean. Essentially, 
tasks are the basis for intent-driven testing. The tasks should 
ideally correspond to semantically meaningful steps when testing AUT, as 
well as align with coherent functionalities of the target application. 
In short, the tasks should be those that a human would want to achieve 
next, given the current testing state.

Generating a viable but also diverse task is crucial, or the exploration risks 
being stuck trying to repeatedly perform impossible, irrelevant, or already 
achieved tasks. We achieve this by combining information from three different sources into a prompt for the planning LLM agent:

\subsubsection{High-level task history}

To continuously generate diverse and consistent tasks, the planner should be 
aware of the past exploration history. Instead of history of GUI 
actions, the planner is provided with textual summaries of $N$ (20 in our 
experiment) most recent tasks, and $M$ 
(five in our experiment) most relevant task knowledge. The historical 
information is inserted into long-term memory by the Reflector, described 
in~\ref{sec:framework-reflector} below, and then retreived and assembled by the 
task retriever, described in~\ref{sec:framework-knowledge-task}.

\subsubsection{Total and visited activities}\label{sec:framework-planner-activities}

\name maintains the number of times each activity has been visited in the spatial memory, and the Widget retriever then includes the list of covered/uncovered activities as a proxy for the exploration progress together with information about all activities as well as the activity of the current state.

\subsubsection{Initial knowledge}\label{sec:framework-planner-initial-knowledge}
\name is designed to be initialised with initial knowledge. In default mode, a profile of a virtual user (persona), the ultimate ``goal'' of the persona, and a sentence denoting the beginning of the exploration, \texttt{[PERSONA] started [APP\_NAME]} is used.

In our experiment, we include the following goal description to facilitate overall diversity: \texttt{[PERSONA]'s ultimate goal is to visit as many pages as possible and try their core functionalities}. 
However, one can customise the goal to adapt to different exploration and testing modes as well. Assuming a social messenger app, the human tester invoking \name could provide a goal such as: \lstinline{[PERSONA]'s ultimate goal is to check whether the app supports interactions between multiple users}.

We additionally guide the LLM implementing the Planner agent to further consider several generally desirable properties when generating a new task: diversity (the task should cover new functionality), realism (the task should be possible on the app), difficulty (the task should be feasible in the fixed action length limit), and importance (the task using core and basic functions of the app should be prioritised).

\subsection{Actor and Observer}\label{sec:framework-actor}
The short-term working memory stores the current task's execution 
history, and is cleared after each planning step to register the newly 
planned task. Both the Actor and Observer access it to retrieve context, and save 
their actions and observations.

\subsubsection{Function-call based action selection}
Actor chooses the appropriate next action to achieve a given task. To reduce 
the number of tokens in LLM prompts, the Actor is given a set of action types
(such as ``touch'' or ``set\_text'') as well as a list of widgets in the current 
screen, instead of combinations of them. Actor subsequently selects an Android 
action type, such as ``touch''  or ``set\_text'', as well as a target widget to 
apply the action to. The prior actions, most recent observation (response) of 
the application, and the critique of recent progress (explained below), are 
also provided to be considered when recommending the next action to try.

\subsubsection{Additional action types}
Actor also supports three types of actions not directly derived from the widgets of the current page: ``wait", ``back", and ``end\_task". 
Mobile testing tools have struggled with detecting loading screens, often using prolonged wait times after each action. The loading screen's presence can be identified by checking for loading messages or icon resource identifiers, and we discovered that the LLMs we used can quite effectively detect loading screens and decide to wait.
So, instead of a fixed long wait, we let the Actor decide when to wait. 
The ``back" action is for navigating back, and the ``end\_task" action allows the Actor to conclude the task before the fixed max action limit (13 in our experiments).

\subsubsection{Observing and summarising the outcome(s)}
The state of GUI may change after taking an action. \name updates its 
perception of a screen with a structured textual representation (JSON). 
However, for the Actor to capture the current task context, it needs to be 
informed about the outcome of the previous action. 
We use a separate Observer agent to summarise the pertinent outcome of an 
action based on a diff of the prior and updated GUI states represented as 
multi-line strings. This is because representing both the prior and updated 
state would lead to long prompts that may confuse the LLM.

\subsubsection{Self-critique}
The Actor may not always choose the desired action. 
Once Actor starts down a wrong path by initiating an undesirable action, it 
becomes challenging to ``escape'' from that incorrect exploration trajectory. 
Therefore, besides offering action results as observations, we incorporate an 
additional element called ``self-critique'' into the Actor of \name. 
Periodically (after every three actions in the experiments), the self-critique 
element generates feedback based on the task execution history up to that point 
and the current GUI state description. 
This involves a separate prompt, which explicitly asks for both a review of the 
task execution history and, if the Actor appears to be struggling, a suggested 
workaround plan. The prompt is sent to a more advanced model, GPT-4, while the 
``main'' conversation querying the next action is handled by GPT-3.5.
Consequently, the generated critique is injected to the Actor's prompting 
context for selecting the next action.

\subsection{Task Reflector}\label{sec:framework-reflector}
Once a task execution round finishes, either by the Actor calling the ``end\_task'' function 
or reaching the maximum action length limit, Reflector is activated to
reflect on and create a concise description of the results of trying to perform the task (binary label indicating task success or failure as well).
The input to this process is the entire task execution history including the
self-critique and all observations from the working memory, the current GUI state, and the ultimate goal (from task planning). We instruct the Reflector to ``derive memorable reflections to help planning next tasks and to be more 
effective to achieve the ultimate goal''. We found that this elaborate reflection
process can help avoid that the overall system ``forgets'' useful knowledge acquired during task execution, given that individual agents summarise their knowledge.
We also found that having different agents focused on specific tasks also helps avoid that involved LLM instances drifts from their purpose, i.e. starts hallucinating or straying from their intended function.

\subsection{Memory Retrieval Modules}\label{sec:framework-knowledge}

\subsubsection{Task Retriever}\label{sec:framework-knowledge-task}
Long-term memory contains a history of performed task, i.e. task-specific 
knowledge as well as reflections on whether the task succeeded (indicates this 
task is supported by the app) or not. \name uses the 
textualised GUI state (basically a concatenation of widget properties) 
captured at task initiation as key\slash query for storing\slash retrieving task 
knowledge. This also allows the Planner to obtain task knowledge derived from past
task executions from similar GUI states, helped by the task retriever.

\subsubsection{Widget Retriever}\label{sec:framework-knowledge-widget}
In manual GUI exploration, a tester accumulates knowledge about the AUT by 
interacting with widgets. Similarly, we introduce a memory type akin to human 
spatial memory, which lets the agent ``remember" observations after interacting 
with a widget and to recall them for future interactions. To achieve this, the 
Observer stores the observations not only to the ``volatile'' working memory, 
but also to a specific widget knowledge (spatial) memory. Each widget is assigned with a 
signature containing the activity it is contained in, textual 
properties such as \texttt{resource\_id}, \texttt{content\_description}, and \texttt{text}\footnote{The text property of an editable textfield is omitted; 
if a widget does not contain any stable textual properties, we use its 
``bounds'' property instead, which corresponds to the position and size of the 
widget.}; the signature of the target widget is used
as a primary key to store and retrieve the observation. 
Additionally, the embedding of the GUI state 
serves as a secondary key in order to fetch observations from the actions 
performed from a similar state to the current one. To limit the amount
of information that goes into the prompts the widget retriever
summarises the $N$ most relevant observations (5 in our experiments), per widget.
This summary is then added to the widget dictionary in the JSON-formatted GUI state, 
under the property \texttt{widget\_role\_inference}.

\section{An Illustrative Example}
\label{sec:example}
In this section, we illustrate the execution of \name with an example. We use 
the ``AnkiDroid'' as the target application, a flashcard 
app for memorising various information. We inject a basic profile of a 
specific user persona, named ``Jade Green'' into the prompt of the Planner and 
Actor model, with account credentials associated with the app.

\subsection{Planning a new task}
The exploration begins with the task planning phase by Planner, where it sets a plan based on its goal and desirable properties of the task.
The answer format is enforced by a template in the prompt; generated answer contains a reasoning step, a suggested task, and an end condition of the task. The end condition is further provided to Actor with the task itself to help Actor decide whether the task is accomplished or not. The following are the actual reasoning steps produced by \name, for the task ``Create a new flashcard'':

\begin{lstlisting}[basicstyle=\scriptsize\ttfamily]
Reasoning about Jade Green's new task: Considering the realism, importance, diversity, and difficulty of the tasks, Jade Green should continue adding flashcards to the newly created "My Deck". This is a realistic and important task, as it is a basic function of the AnkiDroid app and it also helps Jade Green to diversify their activities within the app. This task is not too difficult, as Jade Green has already started this process and is familiar with the NoteEditor page.
Jade Green's next task: Create a new flashcard in the "My Deck" deck with the question "What is the capital city of France?" and the answer "Paris".
End condition of Jade Green's next task: The task is known to be completed when a new flashcard with [...truncated...] is successfully added to the "My Deck" deck.
\end{lstlisting}

\subsection{Executing the task}
Once a task is registered, Actor generates a sequence of GUI actions to 
achieve the task. The taken action and the observation, stored in the working 
memory, are again tied together as a virtual conversation thread between the 
app user and the agent. The following example is the excerpt from such 
a conversation thread. The texts for User are generated based on the action results from the Observer, whereas the ``virtual'' responses of Assistant are the stringified GUI actions previously performed.
On the last User message, the current screen description, with the widget knowledge is appended.

\begin{lstlisting}[basicstyle=\scriptsize\ttfamily, escapechar=|]
* User: My name is Jade Green and I am using an application named AnkiDroid to accomplish the following task: Create a new flashcard in the "My Deck" deck with [...truncated...] What should be the first action?

* Assistant: Fill a textfield that has content_desc "Front" with "What is the capital city of France?"

* User: I performed the action you suggested. What should be the next action?

* Assistant: Fill a textfield that has content_desc "Back" with "Paris"

* User: I performed the action, and as a result, the textfield that had the content_desc "Back" was filled with [...truncated...] This time, I'll give you the full content of the current page as follows: 
        |\color{teal}\{|
            |\color{teal}\textbf{page\_name:} NoteEditor,|
            |\color{teal}\textbf{page\_visit\_count:} 11,|
            |\color{teal}\textbf{children:} [|
            |\color{teal}...|
                |\color{teal}\textbf{ID:} 11,|
                |\color{teal}\textbf{widget\_type:} TextView,|
                |\color{teal}\textbf{content\_description:} Save,|
                |\color{teal}\textbf{possible\_action\_types:} [touch, long\_touch],|
                |\color{teal}\textbf{num\_prev\_actions:} 2,|
                |\color{teal}\textbf{widget\_role\_inference:} The widget allows the user to| 
                    |\color{teal}save their inputs and add new cards, possibly in a|
                    |\color{teal}note-taking or flashcard application.|
        [...truncated...]
        select the next action or end the task by calling one of the given function that corresponds to a specific action.
\end{lstlisting}

Given the conversation thread, the model responds by choosing the next action by calling the corresponding function. The chosen action, touching the ``Save" button, is then converted to an actual GUI event and relayed to the app.
\begin{lstlisting}[basicstyle=\scriptsize\ttfamily]
{
    "name": "touch",
    "arguments": {"target_widget_ID": 11}
}
\end{lstlisting}      

\subsection{Criticising the task execution}

The critique component periodically activates and generates feedback from the 
full task execution history as well as the current GUI state description. As 
other components, the generation of critique is enforced to follow a certain 
template of reasoning step. We demonstrate an actual critique example, pointing 
out that nothing has been added to the desired deck.

\begin{lstlisting}[basicstyle=\scriptsize\ttfamily]
Critique of task execution so far: Jade Green has correctly filled in the question and answer fields, but the flashcard seems to be saved in the "Default" deck instead of the intended "My Deck". This might have happened because Jade Green changed the selected deck from "My Deck" to "Default" after creating the flashcard.
Need a workaround plan?: Yes
Workaround plan for Jade Green: Jade Green needs to ensure that the correct deck ("My Deck") is selected before saving the flashcard.
\end{lstlisting}

\subsection{Reflecting on the task}
At the end of the task execution, the reflector generates a summary of the task 
result, and derives the knowledge from the task execution history. The 
following is the example of the task result and knowledge derivation with the given template for the step-by-step reflection.

\begin{lstlisting}[basicstyle=\scriptsize\ttfamily]
Summary of the task result: Jade Green successfully created a new flashcard in the "My Deck" deck with the question "What is the capital city of France?" [...truncated...]
Task done successfully?: Yes
Reflections on the task:
- Jade Green has learned how to create a new flashcard by filling the 'Front' and 'Back' fields with question and answer respectively and then saving it.
- Jade Green has learned that the app provides a dropdown field to select the deck where the flashcard will be saved.
- Jade Green has learned that the app shows a popup message indicating the number of cards added.
\end{lstlisting}

\subsection{Planning another task with learnt knowledge}

As an exploration continues on and discovers more widgets and activities, 
ideally, the task planning process should benefit from information about the 
past task executions. The subsequent task, ``attaching a photo to the 
flashcard'', illustrated in Figure~\ref{fig-architecture}, is based on 
previously creating a new flashcard. The new task derives new 
knowledge from the observation that the app allows users to add images, audio 
clip, recording audio, and so on, to the flashcard. In future task planning, 
Planner retrieves such knowledge and generate a new task, ``Attach an audio 
clip to the flashcard'' with the following reasoning steps.

\begin{lstlisting}[basicstyle=\scriptsize\ttfamily]
Reasoning about Jade Green's new task: To provide a diverse and realistic task that makes use of the core functionality of the app, Jade Green should try to add an audio clip to a flashcard, which is an important feature of AnkiDroid to enhance learning efficiency. This task is not too difficult as it is similar to the previous task of adding an image to a flashcard.
Jade Green's next task: Add an audio clip to a flashcard.
\end{lstlisting}

\section{Evaluation}
\label{sec:evaluation}

This section describes our experimental setup.

\subsection{Research Questions}
Our evaluation aims to answer the following questions.

\subsubsection{RQ1. Testing Effectiveness}
How does \name compare to existing exploration techniques in exploring diverse functions within a limited time budget?
With RQ1, we aim to assess the diversity and depth of \name's exploration, primarily based on screen coverage.

\subsubsection{RQ2. Usefulness}
How effectively do the tasks generated by \name serve as maintainable testing scenarios, reflecting the supported functionalities of AUTs? With RQ2, we aim to find out whether the tasks generated by \name are useful as valid test scenarios, which can be used for regression testing or further test case generation.

\subsubsection{RQ3. Ablation} How does each component of the agent architecture impact the agent's exploration effectiveness? With RQ3, we aim to assess the contribution of each component of the agent architecture to the overall exploration effectiveness.

\subsubsection{RQ4. Cost} What is the monetary cost of running \name with the latest state-of-the-art large language models? With RQ4, we aim to present the present-day cost of running \name, and provide a view for adopting \name in practice.

\subsection{Experimental Setup}
In this section, we describe our experimental setup.

\subsubsection{Subjects}

Table~\ref{tab-subjects} shows the 15 subject apps we study. We start the app 
selection from the widely used Themis benchmark~\cite{su2021benchmarking}, 
which originally contains 23 open-source Android apps. We are forced to exclude 
eight apps due to deprecated servers or APIs, three apps whose functionalities 
depend heavily on remote servers and are not easily resettable, one app that 
crashes on startup, and another that has only a single activity. We selected five additional apps from FDroid~\cite{fdroid} to broaden the range of our subject app categories.

\begin{table*}[ht]
\caption{Android applications used in \name's evaluation.}
\center{\scalebox{0.8}{
\renewcommand*{\arraystretch}{0.95}
\begin{tabular}{lrllr||lrllr}
\toprule
App Name     & App ID & From& Category& \# of Activity& App Name& App ID& From& Category& \# of Activity \\
\midrule
ActivityDiary& AD & Themis& Personal Diary & 10& openlauncher     & OP & Themis & App Launcher       & 7  \\ 
AnkiDroid    & AK & Themis& Card Learning  & 22& osmeditor4android& OM & Themis & Map                & 18 \\ 
AntennaPod   & AN & Themis& Podcast Manager& 10& MaterialFB       & MF & F-Droid& Social             & 4 \\  
Markor       & MK & Themis& Text Editor    & 9 & collect          & CL & F-Droid& Form Data Collector& 37 \\ 
Omni-Notes   & ON & Themis& Notebook       & 12& APhotoManager    & AP & F-Droid& Photo Manager      & 9  \\ 
Phonograph   & PG & Themis& Music Player   & 12& MyExpenses       & ME & F-Droid& Expense Tracking   & 40 \\ 
Scarlet-Notes& SN & Themis& Notebook       & 8 & OpenTracks       & OT & F-Droid& Sports \& Health   & 24 \\
commons      & CM & Themis& Wikimedia      & 17&                  &    &        &                    &    \\ 
\bottomrule
\end{tabular}}}
\label{tab-subjects}
\end{table*}

\subsubsection{Metrics}
Our primary metric is screen coverage, with a specific focus on activity 
coverage in Android serving as an indicator for exploration diversity. Activity 
coverage is typically defined by the number of activities accessed during the 
exploration of the AUT. We only take account of internal activities that 
include the package name of the target application, since there can 
be external activities that do not represent any accessible screens within the 
AUT (they typically exist to detect memory leaks or to perform crash reports).

While activity coverage is widely used and effective in evaluating the 
``breadth" of exploration, it doesn't necessarily capture the desired ``depth" 
of the exploration. For instance, an exploration technique might navigate to a 
specific activity, it may also return to the previous one without any additional 
interaction. To further evaluate if the test cases generated by each technique 
encompass the target app's comprehensive functionality, we employ the concept 
of ``feature coverage''. This represents the fraction of functional features 
covered by test cases, as delineated in the taxonomy suggested by Coppola et 
al.~\cite{coppola2022taxonomy}. Given that we do not have precise 
specifications for the subject apps, we categorise all discerned functional 
features of each app identified by all comparison target techniques, until the 
consensus of three authors. We then report the number of features covered 
by each technique.

\subsubsection{Baselines}

We compare \name with the following four baselines described below: %
\begin{itemize}
    \item \textbf{Monkey}~\cite{monkey}: Monkey is a widely used random Android GUI exploration tool for Android.
    \item \textbf{DroidBot}~\cite{li2017droidbot}: DroidBot is a systematic input generation tool for Android GUI exploration.
    \item \textbf{Humanoid}~\cite{li2019humanoid}: Humanoid incorporates a deep neural network model trained using real-world human interactions and produces a sequence of GUI actions.
    \item \textbf{GPTDroid}~\cite{liu2023chatting}: GPTDroid interacts with an LLM in a chat-like fashion to produce a series of GUI actions.
\end{itemize}

Note that, since GPTDroid does not provide a replication package, we 
reimplemented it based on the description in the paper. However, one of their 
component is a distinct local language model that converts the natural-language 
LLM response into a GUI event. The construction of the model requires a large 
amount of labelled data, which we could not replicate in our experimental 
context. Therefore, we replaced this component with a function call-based 
action selector, the same we employed in implementing Actor in \name. Instead 
of the GPT-3 model mentioned in the original paper, we used the GPT-3.5 (16K 
context) model. We refer to this reimplemented version as GPTDroid in the rest 
of the paper. 

For each tool, we allocate a two-hour exploration budget. We set up an emulator 
(Nexus 7, API 25) with 2GB RAM and a 1GB SDCard. Each tool runs on a 
64-bit Ubuntu 20.04 machine with an i7-1075H CPU (12 cores) and 32GB memory.

\subsubsection{Large Language Models}
We use GPT-3.5 model with extended 16K context length 
(\texttt{gpt-3.5-turbo-0613-16k} from OpenAI) for the action selection model 
implementing Actor and Observer of \name. The summarisation process of the widget knowledge retriever, as discussed in Section~\ref{sec:framework-knowledge-widget}, requires a shorter prompting context, so we employ the standard GPT-3.5 model with a 4K context (\texttt{gpt-3.5-turbo-0613}) for this module.
For the components crucial to outcomes, specifically the Planner, Reflector, 
and self-critique module of the Actor agent, we employ the GPT-4 model (\texttt{gpt-4-0613}).

\begin{table}[t]
\caption{Number of Covered Activities per App By Each Technique}
\scalebox{0.76}{
\renewcommand*{\arraystretch}{1.1}
\hskip-0.3cm\begin{tabular}{lrrrrrr}
\toprule
Subjects         & \name      & DroidBot  & GPTDroid  & Humanoid   & Monkey     & Total \\ \midrule
APhotoManager    & \textbf{5} & \textbf{5}& 4         & \textbf{5} & \textbf{5} & 9 \\
ActivityDiary    & \textbf{10}& 3         & 6         & 5          & 5          & 10 \\
AnkiDroid        & \textbf{15}& 14        & 6         & 13         & 13         & 22 \\
AntennaPod       & 4          & 3         & 1         & \textbf{5} & 3          & 10 \\
Markor           & 4          & 4         & 4         & \textbf{5} & \textbf{5} & 9 \\
MaterialFB       & \textbf{3} & 1         & 3         & \textbf{3} & 2          & 4 \\
MyExpenses       & \textbf{15}& 7         & 12        & 7          & 11         & 40 \\
Omni-Notes       & 5          & 3         & 5         & \textbf{6} & 3          & 12 \\
OpenTracks       & \textbf{16}& 7         & 11        & 10         & \textbf{16}& 24 \\
Phonograph       & \textbf{11}& 7         & 6         & 9          & 9          & 12 \\
Scarlet-Notes    & 3          & 3         & \textbf{4}& 3          & 3          & 8 \\
collect          & \textbf{13}& 12        & 2         & 9          & 9          & 37 \\
commons          & \textbf{14}& 11        & 7         & 12         & 5          & 17 \\
openlauncher     & \textbf{6} & 2         & 3         & 3          & 4          & 7 \\
osmeditor4android& 9          & 5         & 6         & \textbf{12}& 8          & 18 \\ \midrule
Total             & \textbf{133} & 87       & 80         & 107          & 101     & 239   \\
\bottomrule
\end{tabular}
}
\label{tab-activity-coverage}
\end{table}

\section{Results}
\label{sec:results}

We present the results of our evaluation in this section.

\subsection{Testing Effectiveness (RQ1)}

Table~\ref{tab-activity-coverage} displays the number of activities covered by \name compared to other testing techniques for each application. On average, \name achieves an activity coverage of 60.7\%, slightly exceeding the best baseline, Humanoid, with an average coverage of 51.4\%. A Wilcoxon signed rank test indicated that the number of activities covered by \name was statistically significantly higher than those covered by Humanoid ($p < 0.045$). For those applications that \name discovers significantly more activities than other baselines, the functionalities supported by the applications are relatively intuitive and follow the common sense. However, \name finds it more difficult to visit more activities compared to other baselines against some specific AUTs: our analysis shows that these apps either have widgets that do not have any textual properties (e.g., the widgets in Scarlet-Notes consist of only icons or images without content descriptions), or has a single view containing distinct interactable subregions (e.g., Google Map view on osmeditor4android).

\begin{figure}[t]
    \centerline{\includegraphics[width=0.45\textwidth]{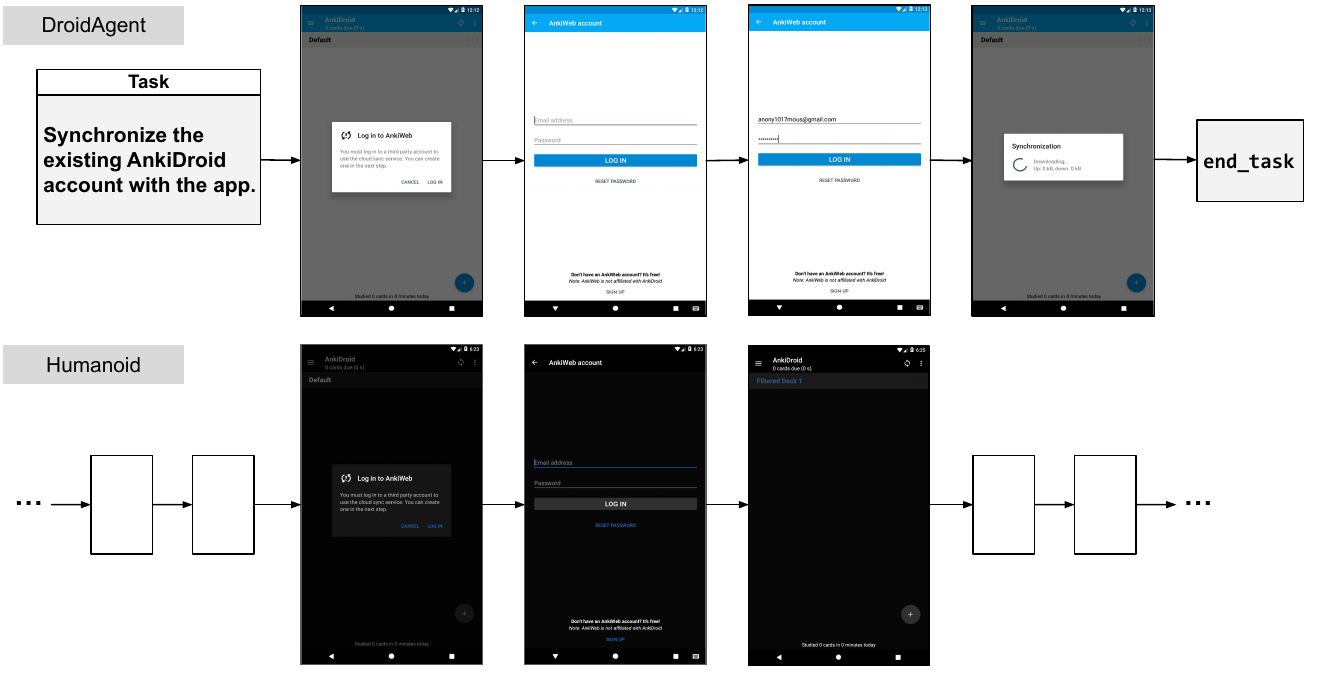}}
    \caption{Example of distinct exploration patterns on MyAccount activities from AnkiDroid application between Humanoid and \name.}
    \label{fig-sync-anki}
\end{figure}

\begin{table}[t]
\newcommand{\crashfound}{\cellcolor{gray!15}$\Circle$}
\caption{Crash Detection Results}
\center{\scalebox{0.8}{
\renewcommand*{\arraystretch}{1.0}
\hskip-0.3cm\begin{tabular}{lrrrrrr}
\toprule
Subjects  & \name & DroidBot & GPTDroid & Humanoid & Monkey\\
\midrule
APhotoManager    & \crashfound& \crashfound& $\times$   & \crashfound& \crashfound  \\
ActivityDiary    & $\times$   & $\times$   & $\times$   & $\times$   & $\times$ \\
AnkiDroid        & $\times$   & $\times$   & $\times$   & $\times$   & $\times$ \\
AntennaPod       & \crashfound& $\times$   & $\times$   & $\times$   & $\times$ \\
Markor           & $\times$   & $\times$   & $\times$   & \crashfound& $\times$ \\
MaterialFB       & $\times$   & $\times$   & $\times$   & $\times$   & \crashfound  \\
MyExpenses       & $\times$   & $\times$   & $\times$   & $\times$   & $\times$         \\
Omni-Notes       & \crashfound& $\times$   & \crashfound& \crashfound& \crashfound \\
OpenTracks       & $\times$   & $\times$   & $\times$   & $\times$   & $\times$  \\
Phonograph       & \crashfound& \crashfound& $\times$   & \crashfound& \crashfound \\
Scarlet-Notes    & $\times$   & $\times$   & $\times$   & $\times$   & $\times$ \\
collect          & $\times$   & $\times$   & $\times$   & $\times$   & $\times$ \\
commons          & \crashfound& \crashfound& $\times$   & $\times$   & $\times$ \\
openlauncher     & $\times$   & $\times$   & $\times$   & $\times$   & $\times$ \\
osmeditor4android& $\times$   & $\times$   & $\times$   & $\times$   & $\times$ \\
\midrule
\textbf{Total}   & \textbf{5} & 3       & 1      & 4       & 4 \\
\bottomrule
\end{tabular}
}}
\label{tab-crash-finding}
\end{table}

\begin{figure}[t]
    \centerline{\includegraphics[width=0.45\textwidth]{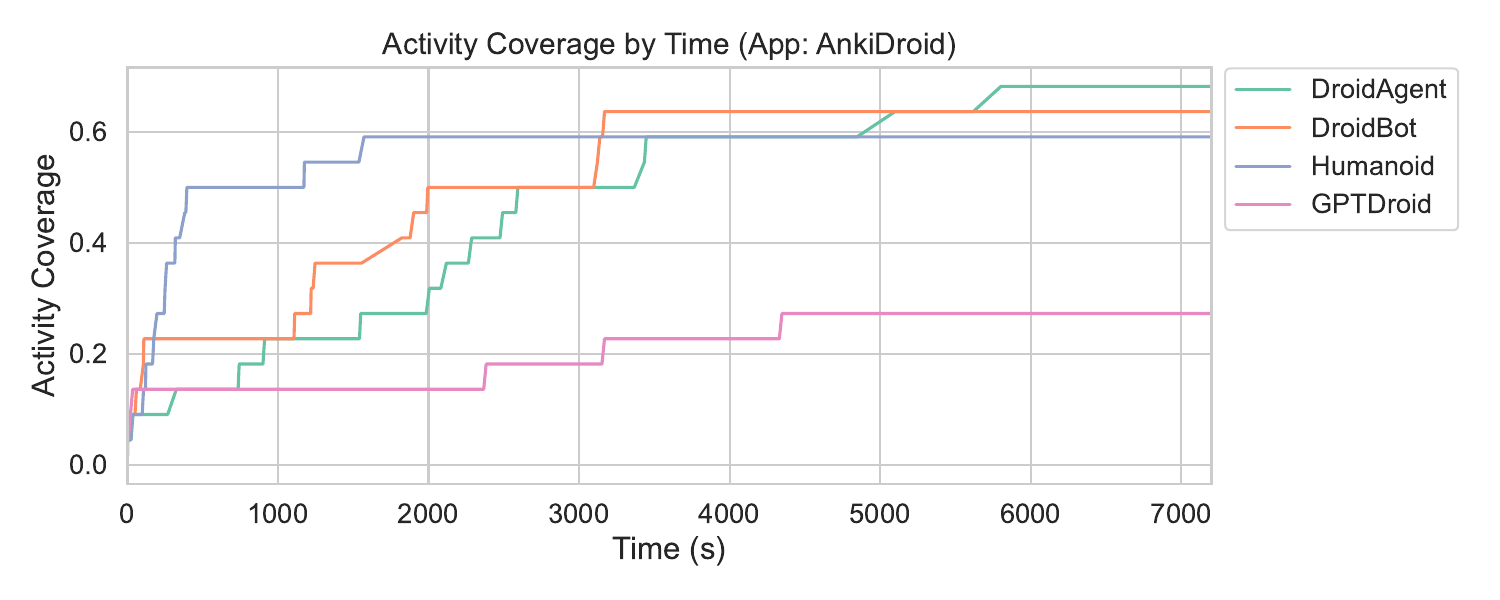}}
    \caption{Activity coverage measured by time for two hours of exploration. (Application name: AnkiDroid)}
    \label{fig-activity-coverage-time}
\end{figure}

Figure~\ref{fig-activity-coverage-time} depicts the change in 
activity coverage over time. We observe the trend on the 
AnkiDroid app as a representative, in which the techniques 
including \name show a similar degree of activity coverage after 
two hours. Humanoid shows a relatively higher growth rate in the 
first 30 minutes compared to others, but it fails to discover 
more activities afterwards. Figure~\ref{fig-sync-anki} 
illustrates one reason for this difference by showing a different 
exploration patterns of \name and baselines on the same activity 
accessed, MyAccount, in the AnkiDroid application. This activity can 
be easily covered by clicking ``Synchronization'' button on the 
main app screen, but the actual synchronisation with the server 
requires logging into the application first. \name succeeds to 
automatically sign into the application with the given profile, 
and wait for the synchronisation to be completed by selecting 
``Wait'' action. On the other hand, Humanoid just triggers 
``BACK" action without any interactions on the activity.

\begin{figure}[t]
    \centerline{\includegraphics[width=0.5\textwidth]{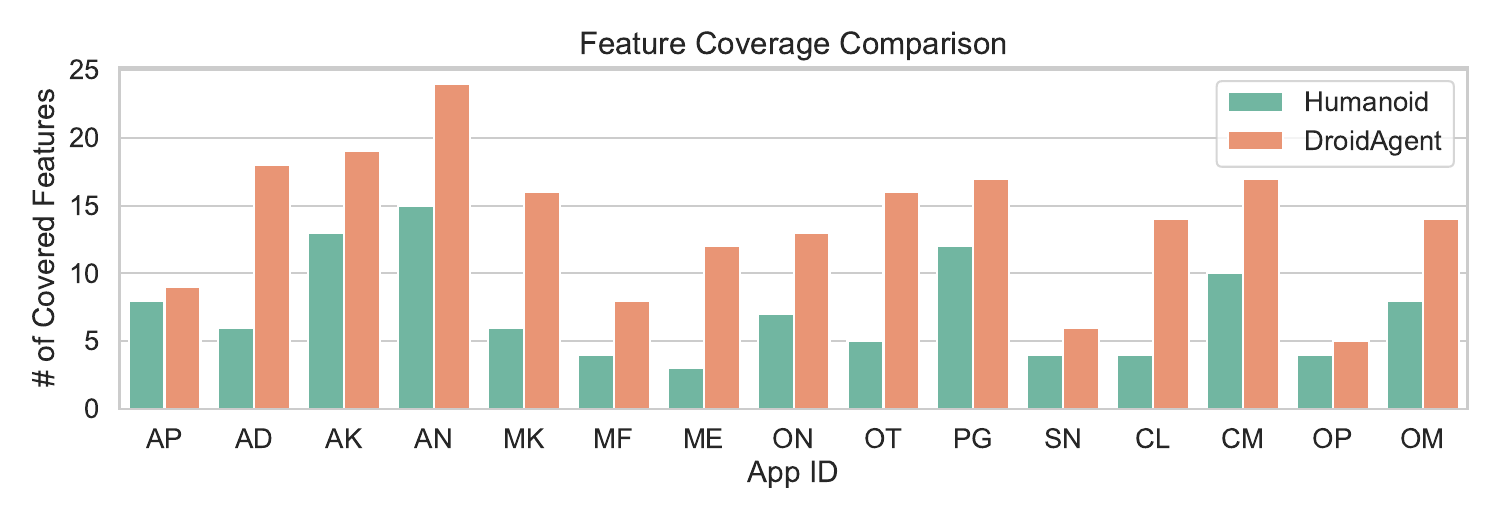}}
    \caption{Comparison of the number of covered features by \name and Humanoid.}
    \label{fig-feature-coverage}
\end{figure}

Figure~\ref{fig-feature-coverage} compares the feature coverage between \name and Humanoid, the best-performing baseline in activity coverage. \name consistently covers a significantly larger number of features than Humanoid (\name average: 13.9, Humanoid average: 7.3) across all subjects; this holds even where Humanoid has higher activity coverage. The result suggests that \name doesn't just navigate the activities, but also engages in meaningful interactions to encompass the features of the AUTs.
A Wilcoxon signed rank test indicated that the number of features covered by \name was statistically significantly higher than those covered by Humanoid ($p < 0.0008$).

Additionally, we report the crashes found by each technique in 
Table~\ref{tab-crash-finding}. \name finds five crashes in total 
(one of them had been reported as a past GitHub issue), while 
Humanoid finds four crashes in total (one of them had been 
reported). It's worth noting that the current goal setting of 
\name, as outlined in 
Section~\ref{sec:framework-planner-initial-knowledge} 
focuses on the \textit{efficient exploration} by covering core 
functionalities within a fixed time budget.

Nevertheless, \name still shows crash finding capability on par 
with other baselines, suggesting its potential as a viable GUI 
testing technique. Further, we argue that the crash (which is 
exposed as an abrupt closing of the app) observed during a 
specific meaningful task, generated by \name, is more readily understood by developers compared to those found during a lengthy yet meaningless GUI exploration. For instance, in the ``commons" app, the application crashes when one attempts to upload a picture and then cancels the process mid-upload. The crash can be presented to a developer along with a task description of image uploading, which we believe aids in more effectively reproducing the issue.

\begin{figure}[t]
    \centerline{\includegraphics[width=0.5\textwidth]{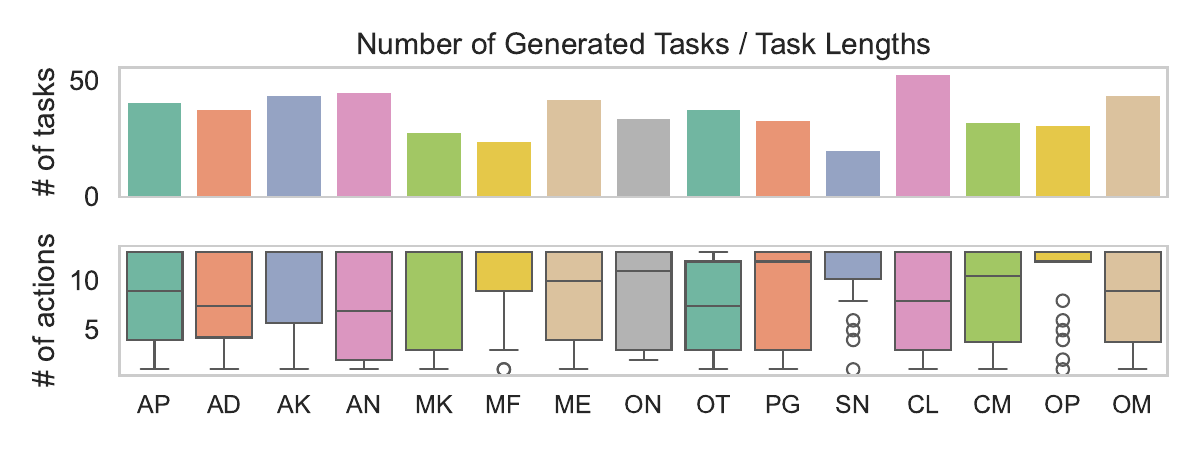}}
    \caption{The number of tasks per application and number of actions per task.}
    \label{fig-num-task}
\end{figure}

\subsection{Usefulness of Generated Testing Scenarios (RQ2)}

We answer RQ2 by assessing the viability and reliability of the 
generated tasks. We first present quantitative 
results for number of tasks, viability, and reliability, and then 
present a qualitative case study of some generated tasks.

\subsubsection{RQ2-1. Task statistics} 

\name planned and executed on average 36 tasks
(standard deviation: 8.8) per AUT. Figure~\ref{fig-num-task} shows the 
distribution of the number of tasks generated per each 
application, and the number of actions (i.e., task length) that 
have been taken to complete each task. Although we set 13 as a 
maximum number of actions per task, the average task lengths vary 
across applications (min: 7.4, max: 11.2, mean: 8.9), as the Actor of \name can end the task earlier.

\begin{figure}[ht]
    \centerline{\includegraphics[width=0.3\textwidth]{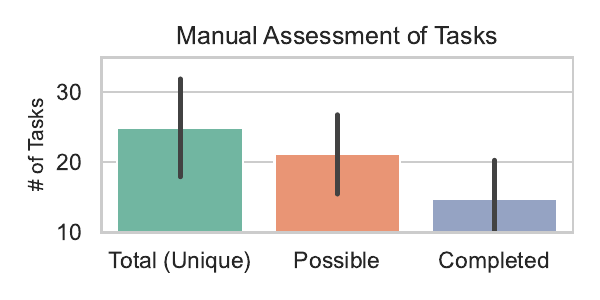}}
    \caption{Manual assessment result of the generated tasks by \name.}
    \label{fig-manual-task-assessment}
\end{figure}

\subsubsection{RQ2-2. Task reliability} \name's Reflector labels 
the task result based on whether the task was successfully completed or not. 
To assess this classification of task results, we have manually 
checked and labeled the completion status of each task, as well as whether the 
tasks are viable with the AUT: a task is viable when it follows 
the supported functionality of the app, and completed when the relevant functionalities of the apps are utilised by the Actor.
Figure~\ref{fig-manual-task-assessment} shows that, among all 
374 unique tasks generated for 15 applications, we deemed 85\% as viable, and 59\% as completed by \name. 
Based on this manual labelling, we report that the Reflector achieves a 
relatively high level of accuracy in task result assessment, with precision of 
0.72, recall of 0.77, and F1 score of 0.74.

\subsubsection{RQ2-3. Case study}
As a case study, we present a couple of tasks generated by \name. 
By producing test sequences in association with ``task'', \name can create complex multi-task scenarios.

\begin{figure}[ht]
    \centerline{\includegraphics[width=0.5\textwidth]{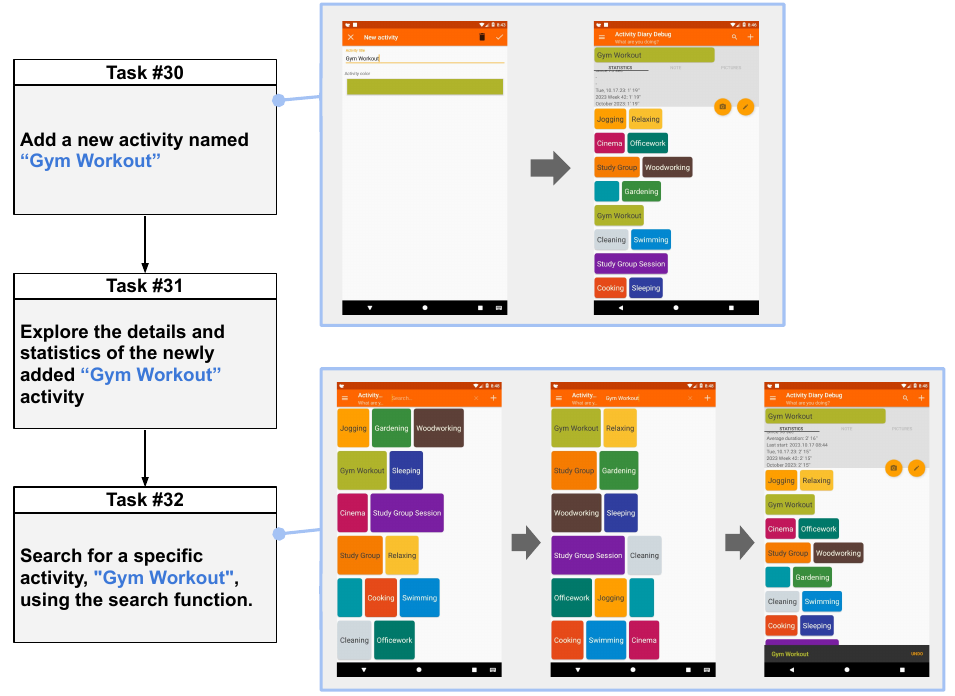}}
    \caption{Example of subsequent tasks consistently reusing generated test data.}
    \label{fig-case-study-1}
\end{figure}

\noindent\textbf{Case 1: Reusing created app data:} We observe that \name is able to reuse app data created during the 
exploration. Figure~\ref{fig-case-study-1} illustrates two 
consecutive tasks in the ``ActivityDiary'' app. \name tries to 
search for a specific activity and successfully inputs a valid 
query, ``Gym Workout'', which is an activity name created during 
a previous task. Subsequently, \name verifies that the targeted 
activity appears on the screen and proceeds to view its details.

Such patterns of \emph{reusing} previously created internal app data 
are commonly observed in our subject applications. For instance, 
in the ``AnkiDroid'' app, tasks like reviewing the flashcard and 
rating the difficulty were conducted \emph{after} creating a new 
flashcard. Compared to \name, baseline techniques often struggle 
to access functionalities that require specific pre-existing app data. For example, they might search for an internal item using a query that is irrelevant and does not yield any search results.

\noindent\textbf{Case 2: Login Automation:} Two of our subjects (commons, MaterialFB) require login steps at startup to access main functionalities. For baseline techniques, as done in prior research, we use login scripts (or setup scripts for simply skipping it). These scripts are either source from the benchmark (Themis) or crafted by the authors (FDroid subjects), aiming at simply bypassing the process. Conversely, \name can autonomously sign into both apps without these scripts when provided with the relevant account credentials in its persona profile, making the sign-in process seamlessly integrates into its exploration routine. \name also exhibits adaptability in handling the app's ``hidden" login features, like the redirected login screen encountered during data synchronisation as previously highlighted (Figure~\ref{fig-sync-anki}). Such scenarios cannot be handled by the login script without adequate prior knowledge of the app's features. 
It's worth noting that even with automated login scripts, the login process might fail due to issues like temporary server errors during login requests. \name can address such flakiness by adaptively retrying the failed action (e.g., re-clicking the login button), making it more resilient.
\subsection{Ablation (RQ3)}

\begin{figure}[t]
    \centerline{\includegraphics[width=0.5\textwidth]{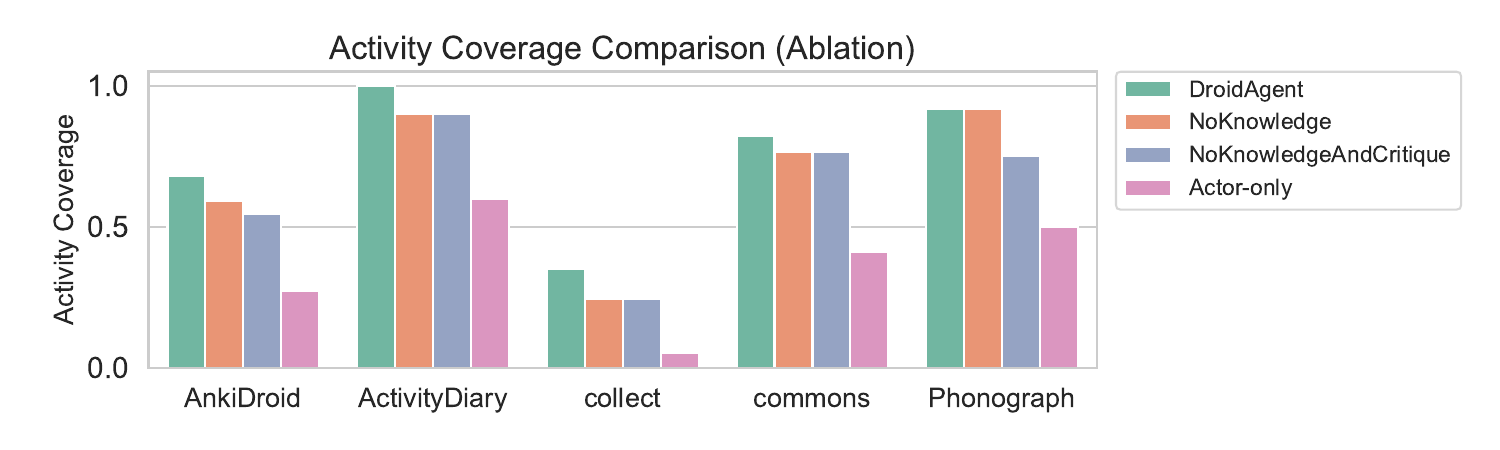}}
    \caption{Average of activity coverage for each ablation setting.}
    \label{fig-ablation-coverage}
\end{figure}

Figure~\ref{fig-ablation-coverage} compares the activity coverage 
of each of \name's ablation settings. We selected a subset of our 
subjects with more than ten activities. 
The ``NoKnowledge'' setting refers to the \name that 
excludes the use of knowledge retrievers. While it incorporates 
\name's Planner and Reflector, the task retriever does 
not supply task knowledge to the Planner. Furthermore, the GUI 
state descriptions given to the Actor and Planner lack 
widget knowledge, as the retriever is disabled.
Similarly, the ``NoKnowledgeAndCritique'' setting refers to the 
\name that additionally excludes the use of the self-critique 
module of the Actor agent. Finally, the ``Actor-only'' setting 
refers to the \name that only utilises the Actor without 
self-critique module. It operates without registered task, 
consistently generating GUI actions based on the current GUI 
state and recent actions.
Both the presence of knowledge retrievers and self-critique 
module seem to positively enhance \name's effectiveness in 
exploring the broader parts of the application.

\begin{figure}[t]
    \centerline{\includegraphics[width=0.5\textwidth]{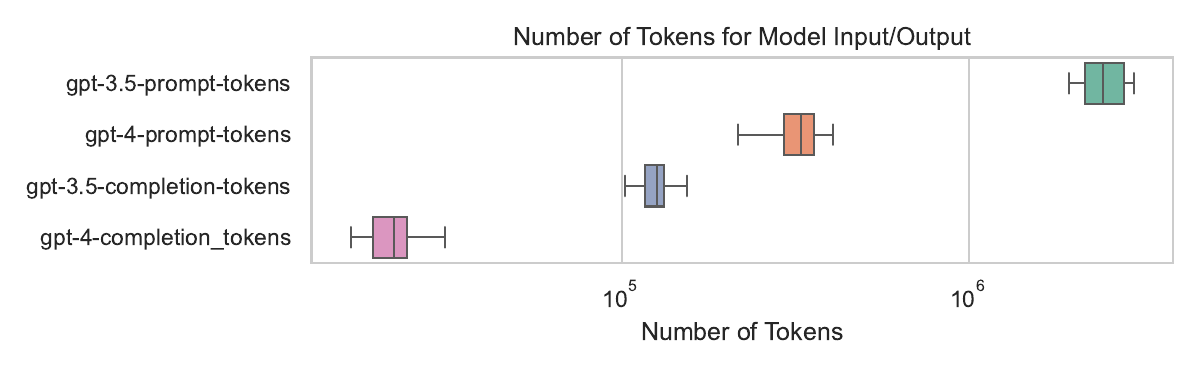}}
    \caption{Number of tokens for GPT-3.5/GPT-4 model input/output (log scale).}
    \label{fig-tokens}
\end{figure}

\subsection{Cost (RQ4)}

Having demonstrated \name's ability to effectively explore app screens, a vital 
question arises: what is the cost of running the agent for app exploration and 
testing? We measured the total number of tokens contained in the prompt and the 
generated output both for GPT-3.5 and GPT-4 models, as shown in 
Figure~\ref{fig-tokens}. The number of tokens for the prompt depend on the 
complexity of GUI layout of each application. Accordingly, the present-time 
cost for running \name on a single application with a two-hour budget ranges 
between \$13 to \$22, summing up the cost from both the GPT-3.5 and GPT-4 
models, averaging \$18.1. Given the trend of decreasing cost per token charged 
by OpenAI, as well as the rapid advancements of open source LLMs, we expect the 
cost of running \name to be reduced and affordable.

\begin{figure}[t]
    \centerline{\includegraphics[width=0.5\textwidth]{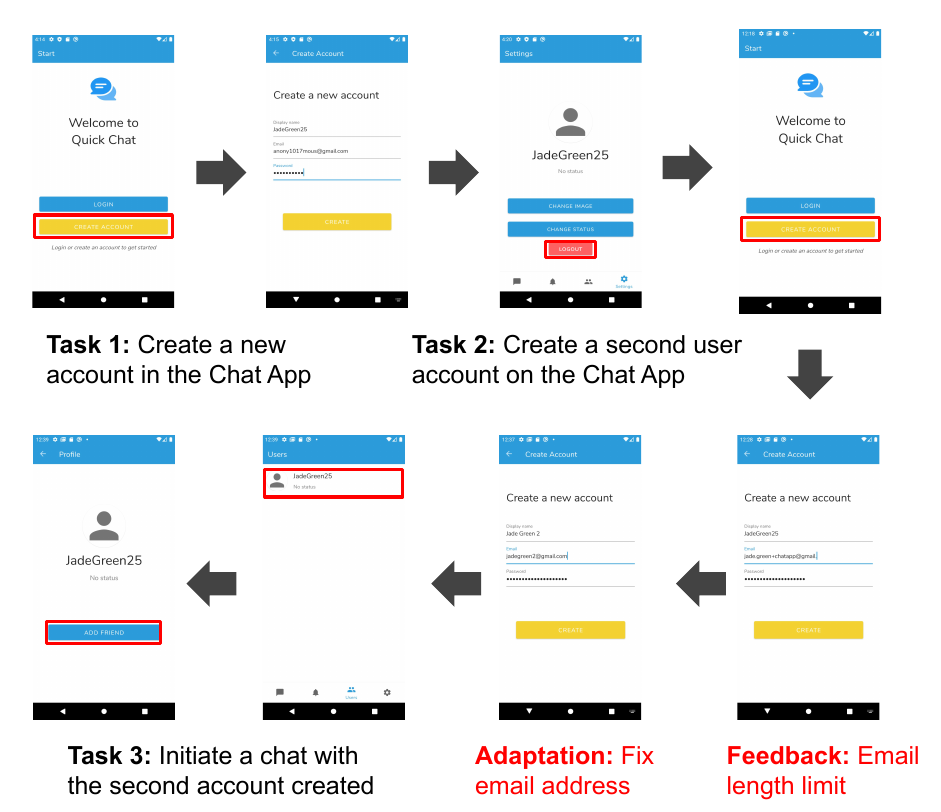}}
    \caption{Example of testing scenarios by \name for creating multiple accounts in a simple chat app.}
    \label{fig:second-account}
\end{figure}

\section{Discussion}

This section decribes a couple of observed behaviour of \name that warrants some discussion and future work.

\subsection{Testing social applications}
So far, testing of social applications that would require multiple accounts has 
been considered out of scope for the existing exploration techniques. We 
demonstrate the potential of applying \name on testing multi-user interactions 
in Figure~\ref{fig:second-account}, which contains testing scenarios generated 
by \name with a custom goal of ``testing multiple user interactions''. The 
first account created follows the persona profile, and the credentials for the 
second account is newly synthesised as a variation of the persona profile. 
Moreover, while creating the second account, \name encounters a truncated email 
address due to the length limit of the textfield, but later it successfully 
works around the issue by using a shorter email address.

\subsection{Testing external use of an mobile application}
A mobile application is not always used in isolation. In fact, it is both 
possible to temporarily navigate out of the app under test and return to the 
app (e.g., selecting a picture from the gallery app, share an app data via 
email), and start the app from the external app (e.g., opening a link from a 
browser). In the former case (temporary navigation to the external app), to 
avoid accidentally being out of the app too long, \name currently imposes a 
fixed interaction limit on external apps and returns to the target app 
automatically. However, we observe some cases that \name prematurely terminated 
essential interactions in the external app due to this limit. Additionally,  
some activities among the subject apps were exclusively triggered by external 
apps, such as the \texttt{WidgetConfiguration} activity, which is only accessed 
by an app launcher. By design, \name is not limited to the target app. 
Broadening \name's scope to test functionalities of AUT across multiple apps 
presents a promising avenue for future exploration.

\section{Threats to Validity}
\label{sec:threats}

\noindent{\textbf{Internal Validity.}} Our study might be affected by the 
inherent randomness associated with LLMs. Given the monetary constraints 
linked to API requests, we could not conduct multiple runs, potentially leaving 
biases. Additionally, one of the baselines, our version of GPTDroid,  includes 
modifications to some of its components. In our implementation, we observed 
that the LLM context limit was reached post ten actions, forcing a reset of the 
preceding conversation prompt, an issue not tackled in the original paper.

\noindent{\textbf{External Validity.}} Our study utilised a relatively limited set of benchmarks as well as underlying LLMs, and therefore may not generalise. We tried to use an existing benchmark of Android apps, Themis~\cite{su2021benchmarking}. Further studies of more apps and other open source LLMs are needed to address this threat.

\section{Conclusion}
\label{sec:conclusion}

We present \name, an autonomous testing agent for Android GUI 
testing. Unlike existing automated GUI testing tools for Android, 
\name sets its own meaningful tasks according to the 
functionalities of the app under test, and subsequently seeks to 
achieve them. Our empirical evaluation of \name against four 
baselines shows that \name is capable of exploring more Android 
activities on average, despite doing so while trying to achieve 
meaningful app specific tasks. \name also exhibits some novel 
behaviour, such as reusing data it created earlier for later 
interactions with the app, or creating multiple accounts to test 
the app. We believe autonomous agents can make significant 
contributions to automation of GUI testing.

\balance
\bibliographystyle{IEEEtran}
\bibliography{ref}

\end{document}